**Title:**

NuQKD: A Modular Quantum Key Distribution Simulation Framework for Engineering Applications

**Authors:**

Konstantinos Gkouliaras,[1, *] Vasileios Theos,[1] Phil Evans,[2] and Stylianos Chatzidakis,[1]
[1]School of Nuclear Engineering, Purdue University, West Lafayette, IN 47907
[2]Oak Ridge National Laboratory, Oak Ridge, TN 37831

**\*Contact Info:**

kgkoulia@purdue.edu

**Abstract:**

A typical experimental Quantum Key Distribution (QKD) implementation requires advanced and costly hardware, not easily available in most university and research lab environments. Testing QKD protocols and evaluating their performance in a particular setup with actual equipment, is usually not possible for researchers. Over the years, this has been a major motivation for the development of simulation frameworks that could effectively model a QKD communication scheme, without the need of an experimental setup. Using this approach, researchers are able to obtain a first insight before proceeding into a practical implementation. Several QKD simulators have been introduced over the recent years. These frameworks have been focused mainly on BB84 as being the first and one of the most explored protocols. Only four out of nine simulators are publicly available. Of these four, only one models equipment imperfections. Currently, there is no simulation framework that includes all of the following capabilities: (i) channel attenuation modelling, (ii) equipment imperfections and their effects on key rates, (iii) estimates of computational time and time elapsed during processes involving the quantum channel, (iv) use of truly random binary sequences for qubits and measurement bases, (v) customization of shared bit fraction, and (vi) open-source availability. In this paper, we present NuQKD, an open-source modular and easy to use QKD simulation framework with all of the above capabilities. NuQKD is a Python script package which simulates the BB84 protocol by establishing a communication scheme between two computer terminals. The code accepts custom input values (number of key iterations, raw key size, attacker interception rate etc.) and provides evaluation of the sifted key length, Quantum Bit Error Rate (QBER), elapsed communication time and more. NuQKD capabilities include simulation of BB84 in both optical fiber and free space quantum channels, modeling of equipment imperfections (source, channel, detector), bit strings from True Random Number Generator, two-terminal implementation, modular design, automated evaluation and export of performance metrics. We expect NuQKD to enable convenient and accurate representation of actual experimental conditions.

## 1. Introduction

An experimental QKD implementation requires advanced and costly hardware, not easily available in most university and research lab environments. Testing QKD protocols and evaluating their performance in a particular setup with actual equipment, is usually not possible for researchers. Over the years, this has been a major motivation for the development of simulation frameworks that could effectively model a QKD communication scheme, without the need of an experimental setup. Using this approach, researchers would be able to obtain a first insight before proceeding into a practical implementation. Several QKD simulators have been introduced over the recent years. A comprehensive survey of simulation frameworks is found in [2]. The survey reviews nine simulation tools found in literature. These frameworks have been focused mainly on BB84 as being the first and one of the most explored protocols. Only four out of nine simulators are publicly available. Of these four, only NetSquid [5] models equipment imperfections. However, NetSquid is a quantum network simulator and therefore might not be the optimal tool for modeling point-to-point communications. None of the frameworks have been validated at the same time by analytical, numerical, and experimental measurements [2] .

Among simulation tools not focused on quantum networks, only EnQuad [1] provides an open source code available for testing [2]. EnQuad is a BB84 open-source, publicly available simulator coded in MATLAB. Despite being an admittedly valuable tool designed for modularity, it makes several assumptions that might not closely reflect a real-world QKD system. All protocol sequences are pseudorandom, generated by the default function rand() with the Mersenne twister. EnQuad further assumes an ideal single photon source, which has not become possible as of yet, therefore ignoring the side effects introduced by the usage of weak coherent laser pulses (e.g., multiple photons, time delays). When it comes to photon detection, dead time, dark counts and detection efficiency have not yet been implemented into the software. Currently, there is no simulation framework that includes all of the following capabilities: (i) transmission distance of each link and channel attenuation, (ii) equipment imperfections and their effects on key rates, (iii) estimates of computational time and time elapsed during processes involving the quantum channel, (iv) use of truly random binary sequences for data and measurement bases, (v) customization for the fraction of shared bits, and (vi) open-source availability.

In this paper, we present NuQKD, an open-source modular and easy to use QKD simulation framework. NuQKD is a Python script package which simulates the BB84 protocol through a communication scheme between two computer terminals. The two terminals represent communication between a facility, located potentially in a remote location, and an operations center intended to monitor and semi-autonomously operate the facility. The code accepts custom input values (number of key iterations, raw key size, attacker interception rate) and provides evaluation of the sifted key length, Quantum Bit Error Rate (QBER), elapsed communication time and more. NuQKD capabilities include simulation of BB84 in both optical fiber and free space quantum channels, modeling of equipment imperfections (source, channel, detector), bit strings from True Random Number Generator, two-terminal implementation in LAN network, or single-device execution, modular design approach, advanced customization of multiple input parameters, and automated evaluation and export of various performance metrics. The code was extensively benchmarked against analytical, numerical and experimental measurements. The description of the key functionalities and the obtained results are analytically discussed in the following sections.

## 2. Overview of QKD

The main concept behind QKD as an unconditionally secure scheme is relatively straightforward: when non-orthogonal quantum states are used to encode data, unlike classical communications, information cannot be accessed or copied without additional information related to the forming of the states. If such thing is attempted, the decoded data would be random. In the case of polarized photons, the additional required information refers to polarization measurement bases [3]. A generic QKD system consists of the transmitter (Alice), the receiver (Bob) and the QKD link between them. It is therefore a point-to-point architecture by default. The link between sender and receiver consists, in the most generic form, of a classical channel and a quantum channel. The quantum channel (optical fiber or free space) transmits encoded qubits in the form of quantum states (e.g., polarized photons). Any attempted interference of an attacker with the quantum channel will lead to disturbing the quantum states, an event which would be detected. The classical channel is responsible for all follow-up communication and is needed to perform the key distillation process and the encryption/decryption. After the key distribution iteration has been completed, both Alice and Bob will have access to an identical secret, truly random key. This generated key can be subsequently used for message encryption, by forming a ciphertext through a symmetric encryption scheme. The encryption scheme can be one-time pad (OTP) or a conventional encryption algorithm, e.g., AES-128.

The first QKD protocol was introduced in 1984 by Bennet and Brassard and belongs to the group of discrete variable protocols [3]. In this scheme, qubits are represented as single photons in a channel that can be an optical fiber or free space [4]. The key is encoded with polarization of single photons using two polarization bases: the rectilinear (0- and 90-degree state) and the diagonal base (+45 and -45 degree state). A single iteration of key distribution in BB84 begins with the preparation of states by Alice. Alice generates two random bit sequences of equal length: the first corresponds to the sent information bits and the second to the bases used to encode them into qubits. The quantum states are subsequently transmitted through the quantum channel. The qubits are measured by Bob, using random measurement bases. Based on his measurements and the decoding bases selected, Bob decodes each qubit and forms a sequence of bits. This iteration is the only contribution of the quantum channel in the algorithm, as the subsequent procedure is carried entirely on the classical channel. Through a procedure called sifting, the communication parties share their used bases over the public channel and discard every bit not measured with the same basis [14]. For the BB84 protocol, the sifted key size will statistically be 50% of the raw key in length. In the parameter estimation stage, additional bits are shared to identify mismatches between the obtained keys for Alice and Bob, while the key length is further diminished. When a part of the sifted key is shared for parameter estimation, only the remaining bits are potentially usable for forming the secret key. For example, a raw key of 100 photons (bits) will reduce to 50 bits after sifting. If half of the sifted key is shared, it will decrease to 25 bits after parameter estimation. Increasing the number of shared bits provides a more accurate estimation for the error rate, but comes at the cost of forming keys shorter in size. This ultimately affects the rate of key generation. The parameter used for error evaluation is the quantum bit error rate, defined as the number of erroneous bits on the sifted key over its length:

$$QBER = \frac{\text{number of erroneous bits}}{\text{length of sifted key}} \quad [1]$$

After QBER estimation, Alice and Bob further communicate through the public channel to eliminate the dissimilarities and obtain a single error-free key. For the reconciliation stage, classical post-processing error correction codes are implemented. Finally, in the privacy amplification stage, the two legitimate parties attempt to further minimize any information that might have been gained by an attacker. The total reduction of the key is expressed by the secret-key rate (SKR). Secret-key rate is an important parameter since it describes the overall performance of the key distribution and is used to evaluate the time capabilities of a QKD scheme. SKR is defined in Equation 2 as:

$$SKR = \frac{\text{final secret key length}}{\text{sifted key length}} \cdot R_{sifted} \quad [2]$$

Where $R_{sifted}$ is the key rate obtained after sifting. The final secret key is less than the sifted key due to the fact that a number of bits must be shared between transmitter and receiver. The length of the key is affected by the presence and intensity of error sources in the communication. The higher the noise in the channel and the higher the attack level, more bits have to be shared to obtain an error-free key. This eventually reduces the secret-key rate. An optimally selected and well characterized channel is necessary to minimize losses that will decrease the performance. The theoretical secret-key fraction is defined in Equation 3 as:

$$k_{th} = h\left(\frac{1}{2} - q_e\right) - h(QBER) \quad [3]$$

Where the right hard side terms in the equation and are calculated using the Shannon binary entropy while $q_e$ describes the level of attack. Unless the channel is noiseless and there are no attacks, the actual secret-key rate is always less than the theoretical secret-key rate [1].

## 3. Overview of NuQKD

NuQKD is a Python script package which simulates the BB84 protocol through a communication scheme between one or two computer terminals. The option for two terminals experimentally simulates communication between two remote parties. The code accepts custom input values (number of key iterations, raw key size, attacker interception rate) and provides evaluation of the sifted key length, Quantum Bit Error Rate (QBER), elapsed communication time and more. NuQKD capabilities include the following:

- Simulation of BB84 in optical fiber and free space quantum channels
- Modeling of equipment imperfections (source, channel, detector)
- Bit strings from True Random Number Generator (TRNG)
- Two terminal implementation in LAN network, or single-device execution
- Modular design for modeling a variety of systems
- Customization of multiple input parameters
- Evaluation and automated export of various performance metrics

NuQKD allows customization of various simulation parameters by the user, making it possible to model a variety of systems. The results are exported in spreadsheet files and figures are saved as images. An overview of the inputs and outputs of the simulation is shown in Figure 1. A description of key functionalities and obtained results are discussed in the following sections.

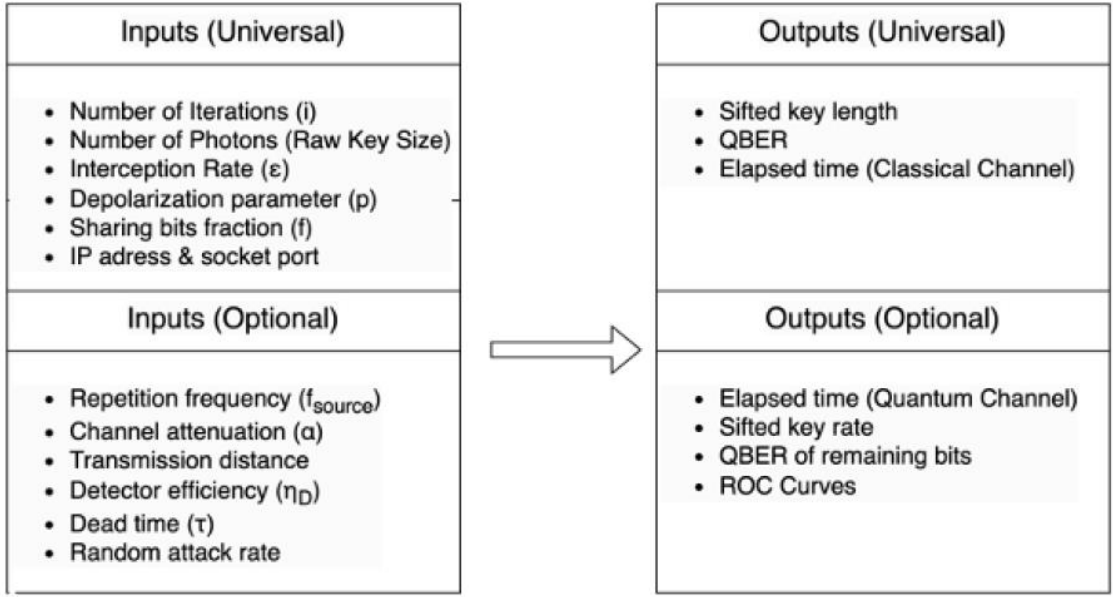

*Figure 1: Overview of NuQKD input and output parameters. NuQKD main function requires a set of standard inputs and generates standard outputs (universal parameters). The user can further enable optional modules associated with dedicated input and output values (optional parameters).*

When the two-terminal option is enabled, NuQKD defines two terminals each corresponding to a sender and a receiver. Each terminal runs a dedicated Python script, and a server-client communication is established in a Local Area Network (LAN). Since the client connects to the server using the TCP/IP address and a dedicated port, it is possible that both terminals run on different computers or even on the same one, if a loopback address is used. NuQKD is designed based on the BB84 protocol. Nevertheless, its modularity allows it to be easily modified to support other discrete-variable protocols. The processes followed by the sender and receiver terminals (Alice and Bob respectively) are schematically demonstrated in Figure 2. The dashed lines signify the optional modules and parameters providing additional functionalities, which can be enabled by the user. The values in blue are the universal outputs, while the ones in green are the optional ones. The modules labeled as Optional 1 are separately shown in Figure 3.

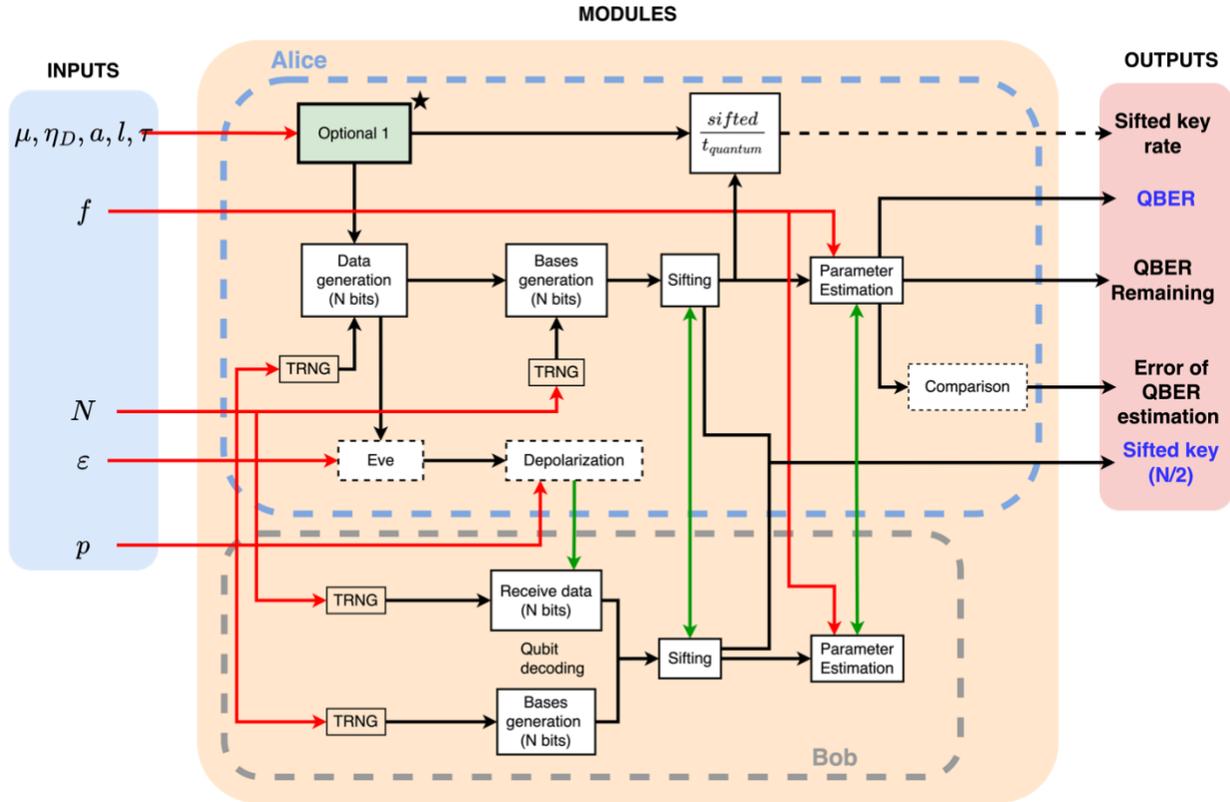

*Figure 2: Block diagram of communication terminals: Sender (Left), Receiver (Right)*

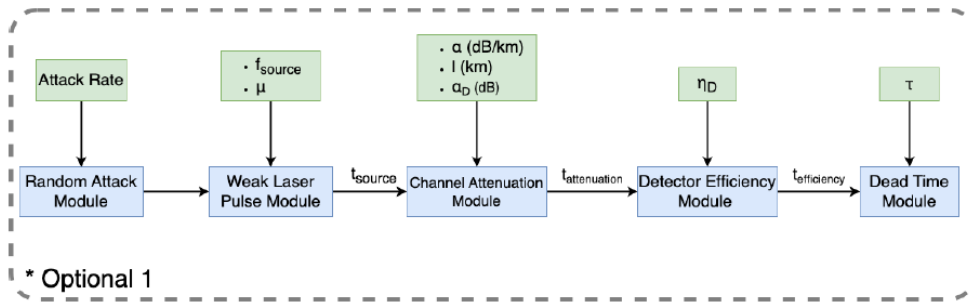

*Figure 3: Block diagram of optional modules*

## 4. NuQKD Model

**Quantum Channel Model**

To achieve a key of length n, Alice needs to form an original sequence of qubits with a length at least equal to N = 4n [6]. To model the qubit states, two data arrays are generated in the first terminal (Alice): a data sequence d, corresponding to a classical bit array and a bases sequence b, corresponding to the selection of bases. Data sequence d is transmitted through the virtual quantum channel. At the same time, the second terminal (Bob) generates a truly random bit string of bases bx′, which is used to decode the received data sequence. In an actual QKD communication scheme, each individual bit of the data sequence transmitted would be correctly decoded by Bob, if the used measurement bases were matched for sender and receiver. In case Bob had chosen the wrong basis to read the qubit (horizontal instead of diagonal basis or vice-versa), the obtained value would not be indicative of the data sequence bit sent, but random. To model this behavior, the bases of Alice/Bob (b and b′) are compared before being exchanged to determine the detection sequence d′. If for any transmitted qubit i ≤ N it is b (i) = b′(i), we assume the detection was correct and we assign d′(i) equal to d(i). Otherwise, the detection decision should yield a random result, thus d′ (i) is assigned 0 or 1 with a 50% probability. b and b′ are sent over the classical channel to Bob and Alice respectively. By comparing these two arrays, communication participants are able to determine which measurements were conducted with a mismatch in selected bases, and thus discard these data points. Since there is a true 50% theoretical probability that Alice and Bob have chosen the same base for each particular qubit transmitted, it is expected that the key sequence after sifting will be on average half the size of the original data sequence.

**Bit sharing**

In the ideal scenario where there is no channel depolarization or any attacker present, the sifted key obtained by the two participants should be identical, and the communication should end there. In practice, any depolarization or interception by an intruder would introduce inhomogeneities between the obtained sifted keys. For this reason, classical signal post-processing techniques are used to correct the error and match the two keys. After both parties have obtained the sifted keys, they choose to share part of the sifted key bits through the classical channel. Therefore, these bits cannot be used for encryption afterwards. The values of the shared bits are compared one to one, and an error ratio is determined. Assuming that the error ratio is approximately the same for the remaining bits, the estimated QBER is calculated by modifying Equation 1 as:

$$QBER_{est} = \frac{number\ of\ erroneous\ bits\ in\ shared\ key}{sifted\ key\ length} \quad [4]$$

The process of creating the shared sequence is illustrated in Figure 4. An important question is what is the minimum shared bits fraction in order to obtain an accurate approximation of the QBER on the remaining bits, which will eventually form the encryption key. As the part of the sifted key shared increases in length, the approximation precision improves. However, this would also reduce the bits available to form the encryption key, which eventually affects the key distribution rate. Therefore, it is important to model this behavior to determine the optimal ratio of shared bits versus the sifted key length.

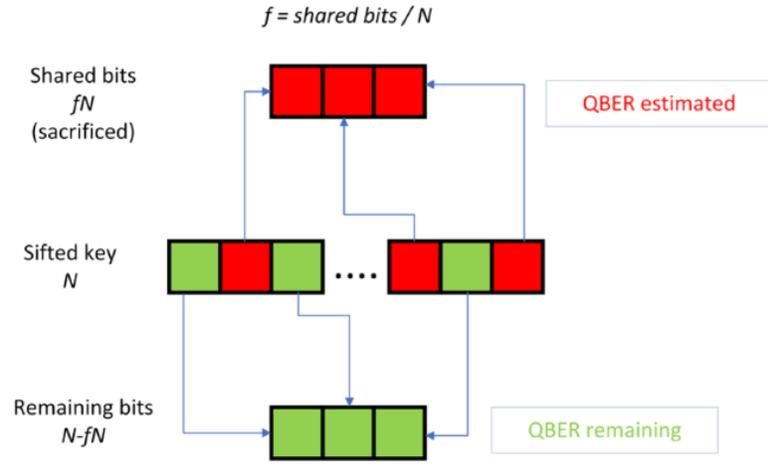

*Figure 4: Procedure for the shared bit string*

**True randomness and connection to quantum sources**

NuQKD simulates several binary sequences, corresponding to the data and bases of the communication parties. A realistic simulation would require these sequences to be independent, and truly random. For a truly random sequence there is no polynomial time algorithm capable of predicting the next bit of the sequence with a probability higher than 50% [11]. NuQKD generates all the bit sequences required for qubit encoding/decoding by fetching true random numbers in real time from a quantum source. Live communication with two different quantum number generator platforms has been integrated in the code in the form of independent key generator functions. The user is also provided with the option of switching between them. The first source option utilized is the Quantum Random Number Generator (QRNG) stream, available online by the Australian National University (ANU) [7]. The ANU QRNG provides a real-time feed of truly random hexadecimal numbers, which we implement in our generated script after converting them to binary. The second available source is the quantum computers available online through IBM's Qiskit platform. Qiskit [12] is an open-source software development kit intended for quantum development.

## 5. Simulation parameters

NuQKD offers customization of multiple simulation parameters. A core functionality requires a set of standard variables to be defined. These inputs are referred to as universal and are discussed in this section.

*Number of photons (Raw keysize):* The user determines the raw key size by specifying the number of photons to be exchanged for each iteration. For BB84 protocol, this corresponds statistically to double the size of the obtained sifted key.

*Number of Iterations (i):* For each key size selected, the user defines the number of key distributions desired. A higher number of iterations provides more accurate statistics.

*Interception Rate (ε):* If this value is greater than zero, a virtual eavesdropper is introduces to simulate an Intercept and Resend Attack. Truly random numbers are generated to obtain the random measurement bases selected by Eve. These bases are used to decode the sent qubits. The result of each photon's measurement is recreated and resent to Bob. A number of pulses transmitted through the quantum channel are interrupted by the attacker. The corresponding qubits are assigned a new value before being resent. If a certain qubit is measured with the correct base by Eve, the original qubit will be resent to Bob. Otherwise, Bob will receive a random qubit as determined by Eve's measurement. It stands for the attack level ($\varepsilon$) and is defined by the user as the fraction of the photons intercepted by the eavesdropper, over the total number of photons sent over the channel for a single key distribution. If set to 1, the attacker attempts to measure every single photon, leading to the maximum introduction of error to the sifted key. Intercept and resend attack simulation can be completely disabled by setting eve to "no" in the central parameter list.

*Depolarization parameter (p):* The depolarization module simulates the effect of a noisy quantum channel. The depolarization parameter p is the total probability a photon is in some way distorted by the phenomenon. The depolarization can either affect the data stream (bit flip), the polarization basis (phase flip) or both (bit and phase flip). These three events are equally probable, with a probability given by p/3 for each [10]. We assume that Eve is located closer to the source than the receiver. This means that if Eve is enabled, channel depolarization takes place -if applicable- after the eavesdropper has interfered with the sent photons. Therefore, channel noise is affecting the photons resent by Eve. Considering the reverse transmission scenario (depolarization preceding eavesdropping) should not alter the overall analysis, given the independence of the effects caused by the two error sources [13]. By combining the depolarization events with eavesdropping, we have formed eight discrete cases of communication interference. Each case has a combinatory effect on the photons Bob actually receives.

*Sharing rate (f):* The sharing rate is a user-defined fraction involved in the classical processing part of the algorithm. It determines the number of bits shared among the parties to estimate the QBER of the key distribution, as a fraction of the size of the sifted key. A higher sharing rate provides more accurate estimation of the error rate, however it comes at the cost of discarding a larger part of the key, eventually affecting the achieved secret key rate.

*IP address and socket port:* The algorithm specifies the port number as the 4-digit number used by the socket to communicate. To connect to the server using the socket port, the client script requires the LAN IP address. In the practical case that the user wants to run the simulation in a single hardware device, the input IP address is set to 127.0.0.1, which is the loopback address.

## 6. Optional Modules

NuQKD provides a number of modules that can be optionally enabled. This makes it possible for the user to even further develop the modules and include additional features. The modules are controlled through altering the central parameter options in the server script. They can be enabled in any combination and the exported data and figures will be automatically adjusted. The optional NuQKD modules and the corresponding input/output variables are summarized in Table I.

*Table I: NuQKD optional modules with dedicated input/outputs*

| Module | Dedicated Inputs | Dedicated Outputs |
|---|---|---|
| Random attacks | Attack rate | - |
| Remaining key | - | QBER remaining |
| Weak pulse laser source | $\mu$, $f_{source}$ | $t_{source}$ |
| Channel attenuation | $\alpha_{ch}$, $l$, $\alpha_{det}$ | $\mu_{effective}$ |
| Detector efficiency | $\eta_D$ | - |
| Dead time | $\tau$ | $t_{dead\ time}$ |

**Random attacks module**

Setting the "eve" option to "yes" and defining an interception rate (single value or range), the user can realistically simulate an intercept and resend attack. The pulses are intercepted at a constant interval for all key distribution iterations. For example, the user could simulate 1000 iterations of 10,000 photons with interception rate 1/2. In this case, all 1000 iterations would be attacked, and the eavesdropper would precisely intercept 500 photons per iteration. In a more realistic communication scenario however, the event of a specific key distribution being attacked could also be random. For example, it would be possible that only a subset of 100 out of the 1000 iterations are attacked and the 100 attacked distribution cycles would be randomly distributed among the total iterations. In order to randomize these events, the random attacks module is introduced. When enabled, the user defines an attack rate corresponding to the fraction of the number of key iterations attacked over the total iterations. By default, this value is set to 1/2. Consequently, only a specific number of iterations will involve the eavesdropper for a particular simulation scenario. It is noteworthy that the indexes of the attacked iterations do not need to be truly random. This is due to the fact that the particular selections are not the result of a physical stochastic process but can rather be attributed to the eavesdropper's strategy. Nevertheless, the simulation is not affected since the QBER detection threshold does not dynamically change but is pre-agreed among the communication parties beforehand. Without knowing whether an attacker is currently present, server and client make a decision to abort or not the communication based only on their QBER estimation. The algorithm stores their decision, as well as the indexes of the iterations attacked. By comparing these values for each time step, True/False Positive and True/False Negative occurrences are calculated and the ROC curves can be constructed.

**Remaining key module**

"Remaining key" is a binary, user-controlled indicator. If turned on, the program performs additional computations without affecting the core functions of the simulations. This module is used to evaluate the agreement of the QBER estimated by the communication parties using the shared key, with the error rate which will appear on the remaining bits of the key. It calculates the level of mismatch on the bits that will be used to form the final secret key. After the shared key is formed on Alice's side, it is shared with Bob through the classical channel to compare. The remaining, non-shared bits of the sifted key will be stored on both Alice's and Bob's side, forming two arrays of "usable keys". The remaining key module will perform an additional round of data exchange to transfer Bob's usable key to Alice and compare the sequences bit by bit, calculating the QBER remaining variable. This variable is stored along the rest in the output spreadsheet. These data are useful for analyzing the efficiency of the error estimation as a function of the shared bit fraction and of the number of sent photons. It should be noted that sharing the remaining bits through the classical channel is a practice that obviously eliminates the motives behind implementing QKD. Consequently, this module is designed to be enabled solely for experimentation purposes (therefore the binary variable is named "research") and is not intended to be a part of the key distribution scheme.

**Weak pulse laser source module**

One limitation of practical QKD is the absence of a deterministic single photon source. Because such source is ideal, weak laser pulse sources are used. Each pulse produced contains a different, probabilistic number of photons. This number is sampled from the Poisson distribution with an average value of photons per pulse $\mu < 1$. Another limitation associated with nonideal photon sources is the introduction of significant time delays. Repeated pulses might be needed to obtain a single photon. Consequently, the time for exchanging the qubits does not solely depend on the optical fiber transmission specifications, but is now dictated by the pulse generation limits of the source. For example, a source with a value of $\mu = 0.1$ photons/pulse will produce zero photons with a 90.4% probability. Equivalently, there is less than 10% chance that at least one photon is produced. If the source has a typical frequency of f = 1 MHz, then it generates 1 million pulses per second. By linearity of the Poisson distribution, in 1 second corresponding to n = 1 million events we would expect on average:

$$\mathbf{E}\left[\sum_{i=1}^{10^6} X_i\right] = \sum_{i=1}^{10^6} \mathbf{E}[X_i] = 10^6 \cdot \mu = 10^5 \; photons \qquad [5]$$

Where $X_i$ is the number of photons from pulse i. An ideal source, on the other hand, would have produced ten times more photons for the same amount of time. The rare occasions when more than two photons are produced from a single pulse will also contribute just a single qubit, in addition to posing a PNS attack risk. For this reason, the number of qubits used to form a secret key will be less than the average number of photons. To simulate the behavior of a weak coherent source, NuQKD has been designed to include the weak laser pulse module. Once this option is enabled, the user can specify both the source frequency and mean number of photons per pulse. For each iteration, the algorithm uses the Poisson distribution to calculate the time required for the photons to be generated. The number of photons needed is specified by the user-defined universal simulation parameters. Since this time interval is expected to be much larger than the transmission time in the fiber, it is approximated that it corresponds approximately to the total time elapsed for quantum channel communications, as given by Equation 6:

$$t_{\text{quantum}} = t_{\text{source}} + t_{\text{transmission}} \cong t_{\text{source}} \qquad [6]$$

In addition to storing the individual values, the statistics of $t_{\text{source}}$ are also calculated over multiple iterations (average, standard deviation) and the respective figures are exported.

**Channel attenuation module**

In an earlier section the channel attenuation was discussed as a characteristic of each channel. In real-world setups, the detector introduces further attenuation to the received optical signal, usually by a standard value $a_{det}$. The unified attenuation coefficient $a$ is then the sum of the channel attenuation (function of the transmission distance l) and of the detector-imposed attenuation, thus:

$$a[dB] = a_{ch}\left[\frac{dB}{km}\right] \cdot l[km] + a_{det}[dB] \qquad [7]$$

Practically speaking, the effect of attenuation to our simulation is extending the time required to obtain a sifted key, eventually lowering the secret key rate. Equivalently, we can model the effect of attenuation by altering the mean number of photons contained in each pulse from the source. The effective mean photon value is a function of the total attenuation coefficient α and the source attenuation μ. In particular:

$$\mu_{\text{effective}} = \mu \cdot 10^{-\frac{a_{ch} \cdot l + a_{det}}{10}} \qquad [8]$$

In case a channel attenuation coefficient greater than zero is specified, along with a nonnegative transmission distance, the algorithm calculates the effective µ before proceeding with the weak laser pulse module, and the simulation follows with the updated parameter.

**Detector efficiency module**

The detector efficiency $\eta_D$ is the percentage of received photons that the detector will actually record. For a non-ideal detector, this value is lower than one. The detector efficiency module can be combined with the weak laser pulse option, or simply with an ideal single photon source, as it calculates the additional time that a formation of an N-sized raw key (photon bunch) requires. For an ideal source, a pseudorandom number from 0 to 1 is generated and compared with the detection probability for each generated photon. If the threshold is not exceeded then the photon is counted, otherwise it is ignored. For a weak laser pulse source, the module performs the above comparison only for the cases when the pulse contains one or more photons. However, the detection efficiency threshold shifts for multiphoton pulses, since only one of the photons needs to be detected in order for the photon to contribute. The detection event is the union of the separate events. Given that the detection (or no detection) of each photon in the pulse is an independent random event with probability $\eta_D$, the detection probability threshold for a pulse containing $N > 1$ photons is:

$$p_{\text{detection}} = P(A_1 \cup A_2 \cup \ldots A_N) = \sum_{i=1}^{N} P(A_i) - P(A_1 \cap \ldots A_N) = N \cdot \eta_D - \eta_D^N \quad [9]$$

Where $A_i$ is the event that the i$^{\text{th}}$ photon of the pulse will be detected. This contributes to the total time delay calculated, since the time required for forming a key of given size will increase due to detection imperfection.

**Dead time module**

Dead time is another optional module. It models the effects of the time interval following detection, during which the detector is unresponsive. Dead time (τ) is a considerable source of time delay affecting the secret key rate. Typical values of τ are around 90 µs. In the simulation, it can be combined with any of the other modules. When enabled, the algorithm creates a time variable for each key distribution, and adds τ microseconds to it every time a successful detection is noted. The sum is added to the delays introduced by the source, channel and detector efficiency effects to provide a total value for the quantum channel time delay, as:

$$t_{quantum} = t_{source} + t_{dead\ time} \quad [10]$$

**7. Benchmarking**

NuQKD was benchmarked against analytical, numerical and experimental data. The benchmarking is discussed in this section.

*Numerical*

To evaluate the code accuracy, we tested NuQKD against the EnQuad simulator (Abdelgawad et al., 2020). NuQKD is run once for each depolarizing parameter (one single key distribution each time) and the QBER is evaluated. The process is performed twice, the first time with no eavesdropping at all (ε = 0) and the second one repeated for a full-scale intercept/resend attack (ε = 1). In each instance, the initial number of photons is set to 10,000. We modified EnQuad to iteratively run for the same p and ε values. The results are plotted on Figure 5a. As it is seen, both NuQKD and EnQuad simulators follow the same trendline, with some deviations due to randomness and due to the fact that there is no averaging of the QBER value over multiple iterations.

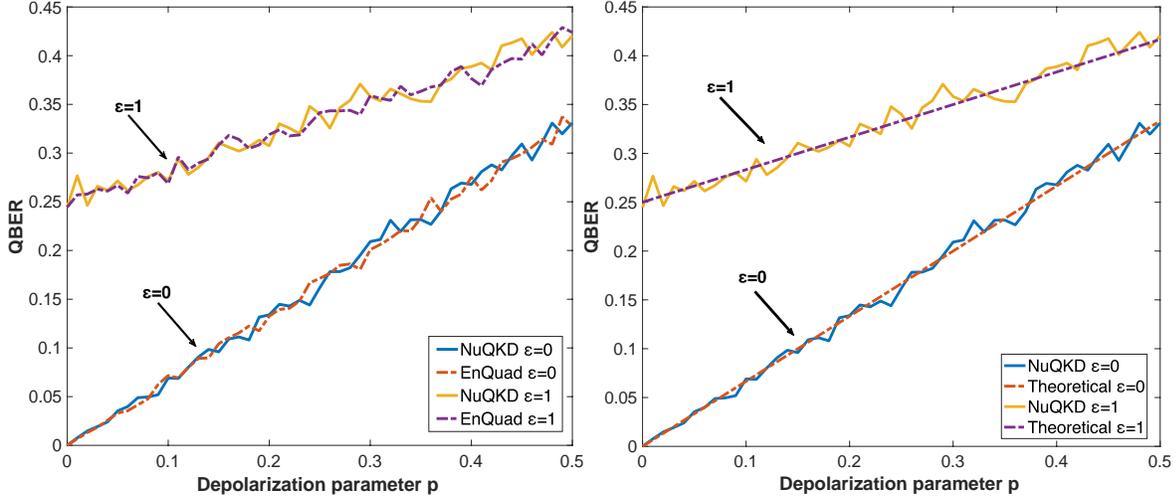

*Figure 5a (left): Benchmarking of NuQKD vs. EnQuad for varying polarization levels (single iteration, photons= 10,000, f = ½, ε = 0 and ε = 1). Figure 5b (right): Benchmarking of NuQKD vs. analytical equation (12).*

*Analytical*

When the only sources of error are channel depolarization and intercept & resend attacks, the theoretical QBER can be described as a function of the depolarization parameter p and attack rate ε:

$$QBER = \frac{\varepsilon}{4} + \frac{p}{3}(2 - \varepsilon) \qquad [12]$$

Same parameters as with numerical benchmarking were used for depolarization, attack rate, sharing ratio and raw key size. The results are shown in Figure 5b. To see how the average QBER varies, NuQKD is run 100 times repeatedly for each depolarization parameter value in the range from 0 to 1. QBER is then averaged over the 100 iterations per p value. The results are plotted in Figure 6 for ε = 0. Each case is simulated for different number of photons per key, ranging from 10 to 10,000. The average QBER is practically indistinguishable from the theoretical result as the photon number increases, while the variance is only visible for shorter key sizes.

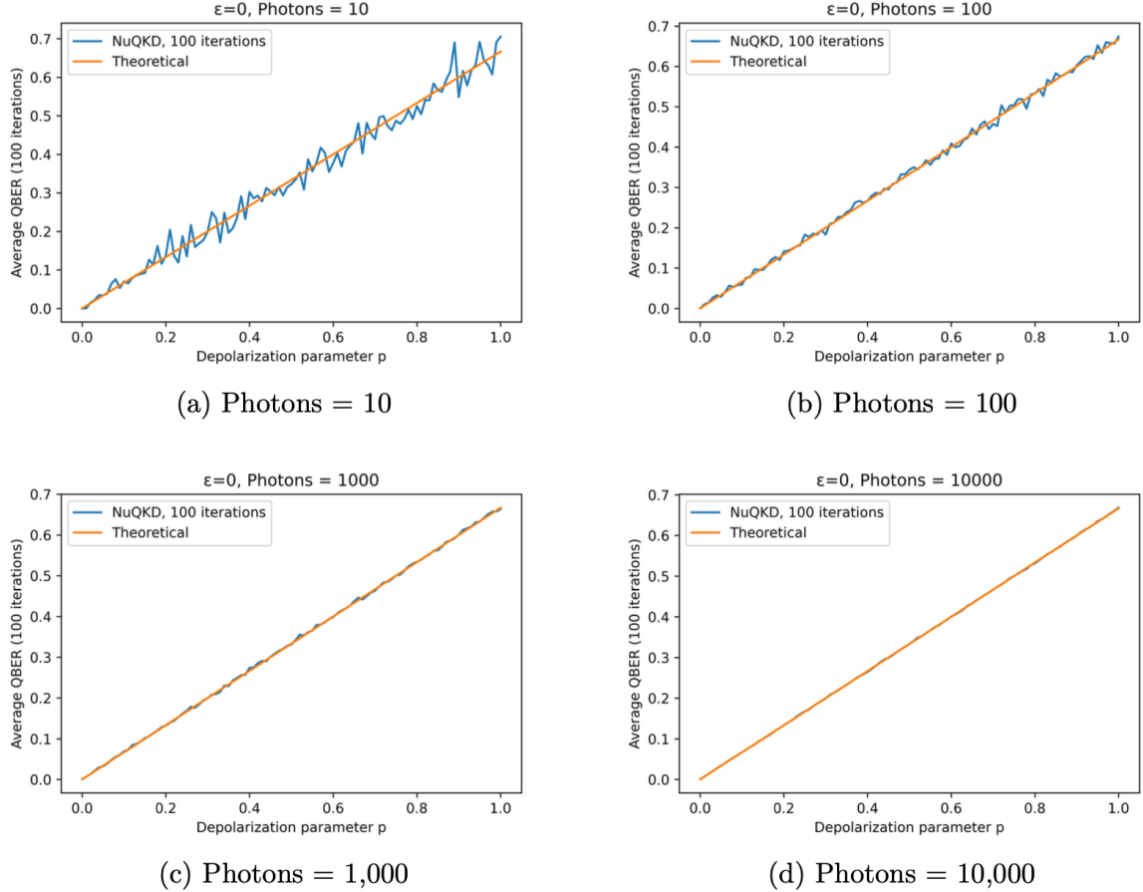

*Figure 6: Average QBER over 100 iterations and various raw key sizes (ε = 0, f = ½).*

*Experimental*

The final stage in the benchmarking procedure was to demonstrate that NuQKD is capable of simulating BB84 under realistic conditions, by taking into account real-world equipment and channel imperfections. Two publicly available practical experiments were selected. The first one is a QKD link over a fiber quantum channel, while the second one is an application where free space is the quantum medium.

The first experiment, performed by Jeong et. al in 2010, investigated the performance of a depolarizing quantum channel, by comparing BB84 and SARG04 protocols [8]. The authors achieved a sifted key rate of approximately 16.5 kbps through a single mode fiber with length of 1.27 km. The channel was characterized with attenuation 3 dB/km at a wavelength of 780 nm. The source was clocked at 1 MHz with a mean photon number was μ = 0.189. Detector efficiency was 0.4 while no dead time was specified.

The second experiment, featured in S. Nauerth's PhD dissertation, involved the experimental implementation of a QKD link between a moving airplane and a ground station [9]. The experiment was intended to generalize free-space optics transmission by including the case of a moving platform. Once again, BB84 protocol using polarization encoding was applied and faint laser pulses were used. For a transmission distance of 20 km, a sifted key rate of 145 bps was achieved, along with a practical error rate of 4.8%. The equipment specifications and channel characteristics for both experiments were recorded and provided as inputs to NuQKD in order to simulate the experiments. These parameters are displayed in Table II.

For a negligible dark count probability, the theoretical sifted key rate can be calculated as a function of the source frequency, the detector efficiency and the effective mean photon number as:

$$R = \frac{1}{2}(1 - e^{-\mu\eta}) \cdot f_{source} \qquad [13]$$

Using Eq. 13, the theoretical sifted key rates for experiments 1 and 2 are calculated equal to 15,4768 kbps and 125.281 bps, respectively.

*Table II: Equipment specifications and channel characteristics of QKD experiments*

| Parameter | Exp 1 | Exp 2 |
|---|---|---|
| Distance | 1.27 km | 20 km |
| Attenuation (channel) | 3 dB/km | 1 dB/km |
| Wavelength | 780 nm SMF | 850 nm FSP |
| Attenuation (Detector) | ~0 | 3 dB |
| Detector efficiency | 0.4 | 0.5 |
| Source Repetition rate | 1 MHz | 1 MHz |
| Pulse µ | µ=0.189±0.001 | µ=0.1 |
| Dead Time | ~0 | 50 ns |

The simulation using NuQKD was run for 1,000 iterations with the Table II values as input parameters. We evaluated the sifted key rate for four different raw key sizes (10, 100, 1,000 and 10,000 photons). The output of NuQKD is very close to the experimental value in all cases. It appears however that using a smaller photon batch provides a more precise average, at the cost of a larger standard deviation (more unstable key rate). As the key size increases, the average sifted key rate slightly falls, but is shown to be consistent with a smaller standard deviation. Regardless of the photon batch size, it appears that the NuQKD outputs agree with the theoretical values. The simulation results are shown in Table III and plotted in Figure 7. The trend of the sifted key mean value and standard deviation for varying photons sent are shown in Figure 9, along with the experimental and theoretical values.

*Table III: NuQKD results from simulating experiments in [8] and [9]*

| Sent photons | Average sifted key rate (bps) ± standard deviation | |
|---|---|---|
| | Exp 1 | Exp 2 |
| 10 | 16,678±7,827 (1.0%) | 141±67 (2.7%) |
| 100 | 15,706±2,178 (4.8%) | 126±17 (13.1%) |
| 1,000 | 15,512±655 (5.9%) | 125±6 (13.7%) |
| 10,000 | 15,494±214 (6.0%) | 125±2 (13.7%) |

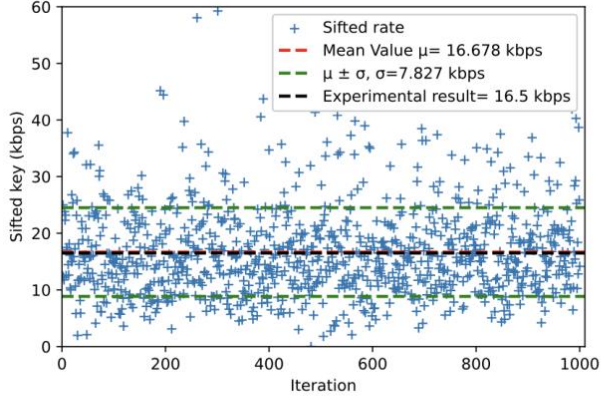
(a) Photons = 10

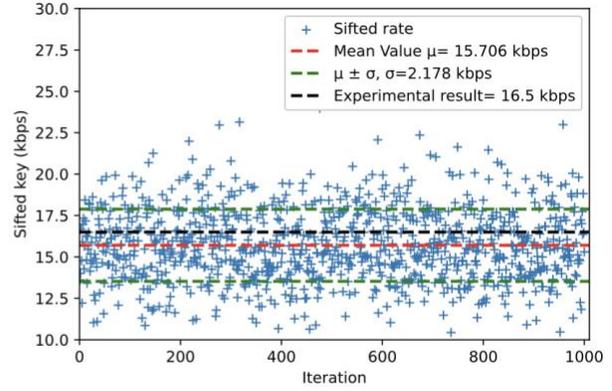
(b) Photons = 100

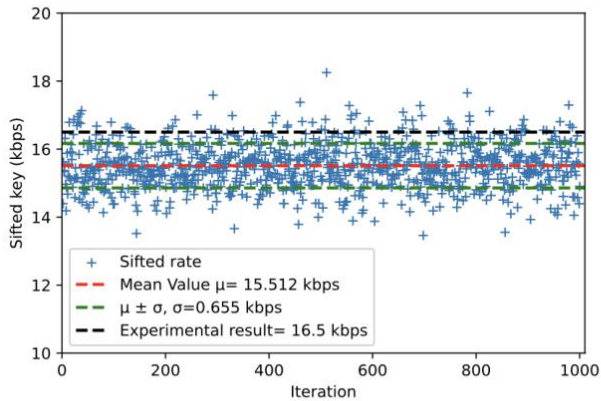
(c) Photons = 1,000

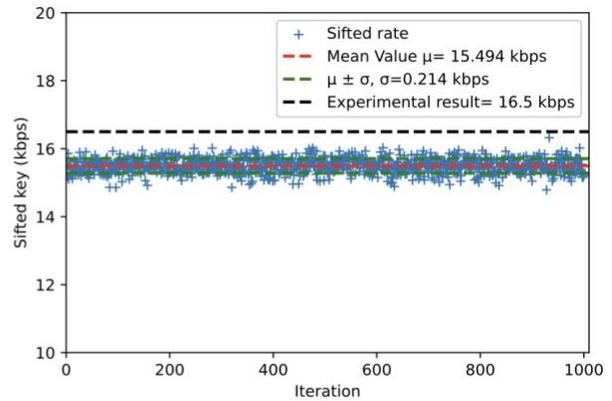
(d) Photons = 10,000

*Figure 7: Sifted key rate evaluation of [8] with NuQKD (SMF, l = 1.27 km, a=3dB/km, λ=780nm, f =1MHz, µ=0.189, η =0.4)*

The simulation of the second experiment also yielded results in good agreement with the experiment, confirming that the algorithm can be used for evaluating the performance of both optical fiber and free-space communication schemes. An average approaching 145 bps is obtained for all raw key sizes. A similar trend as before is displayed for the standard deviation, as the error decreases for larger keys. The sifted key rates are displayed in Table III and they are plotted along with the experimental average on Figure 8. The trend of the sifted key mean value and standard deviation for varying photons sent are shown in Figure 10.

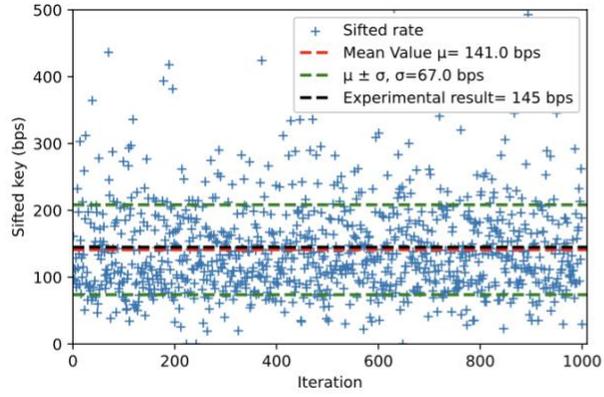
(a) Photons= 10

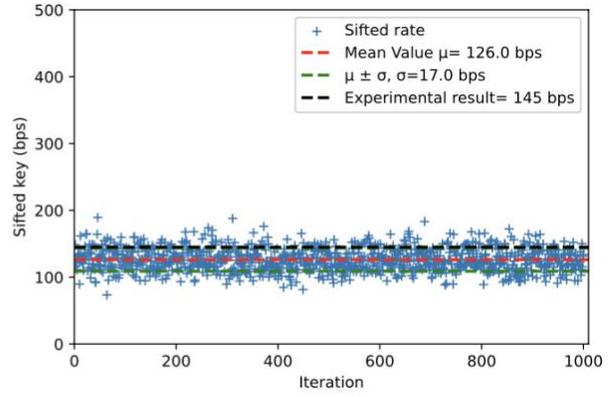
(b) Photons= 100

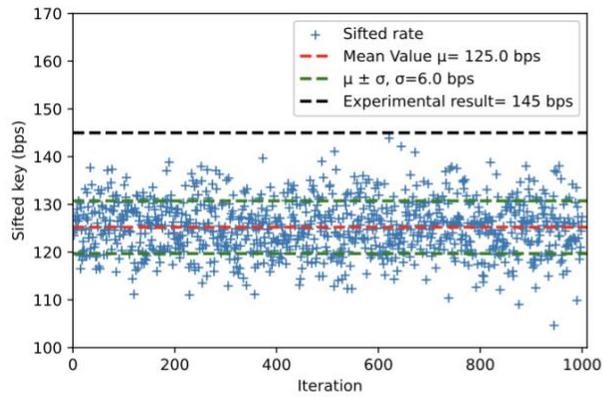
(c) Photons= 1,000

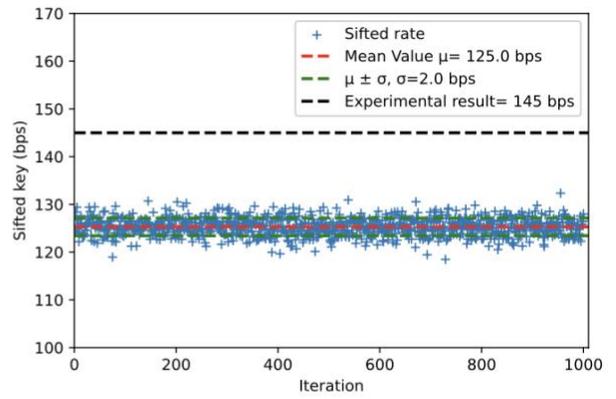
(d) Photons= 10,000

*Figure 8: Sifted key rate evaluation of [9] with NuQKD (FSO, l = 20 km, α=1dB/km, λ=850nm, fsource =1MHz, μ=0.1, ηD =0.5, αD =3dB, τ = 50 ns)*

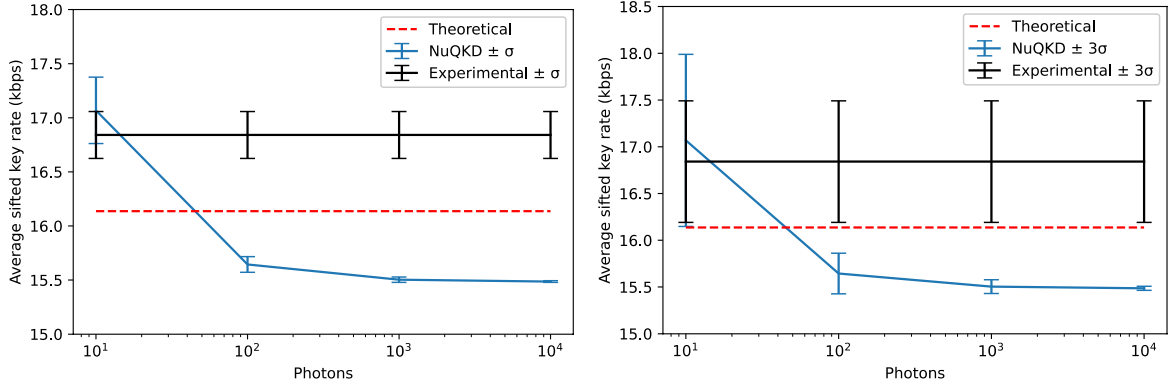

*Figure 9: Mean of the average sifted key rate versus photon number for NuQKD evaluation of [8]. Averaging over 20 simulations, 1,000 iterations each. (SMF, l = 1.27 km, a=3dB/km, λ=850nm, f =1MHz, μ=0.189, η =0.4, )*

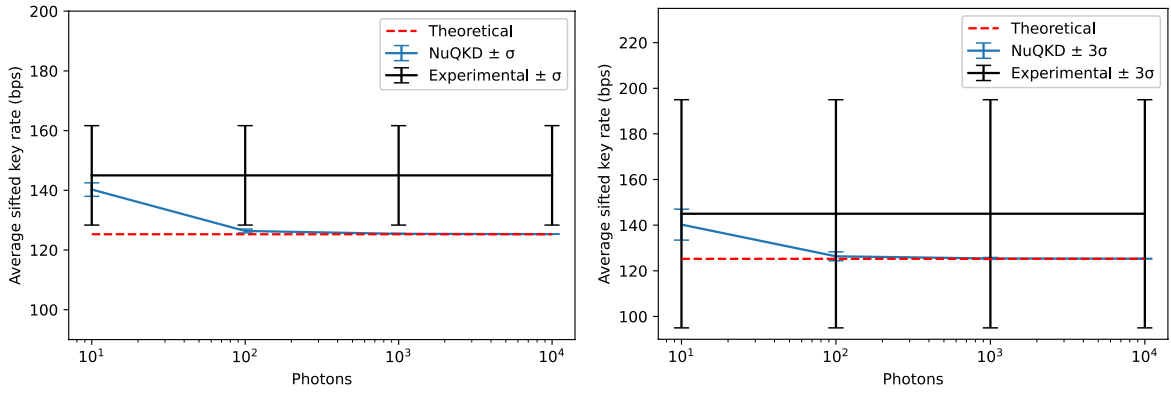

*Figure 10 : Mean of the average sifted key rate versus photon number for NuQKD evaluation of [8]. Averaging over 20 simulations, 1,000 iterations each. (SMF, l = 20  km, a=1dB/km, $α_D$ =3 dB, λ=780nm, f =1MHz, μ=0.1,, η =0.5, τ = 50 ns)*

## 8. Summary

A novel simulation tool (NuQKD) was developed to allow low-level parameter customization and to thoroughly model the actual conditions of a QKD experiment. NuQKD was used extensively to reveal the correlation between different parameter combinations, which significantly influence the performance of a quantum communication scheme. The proposed tool was benchmarked against numerical, analytical and experimental data. For experimental benchmarking, NuQKD was used to replicate the measurements of precedent, real-world experiments. The simulation accuracy was validated for two selected experiments employing both single-mode optical fiber and free-space optics.

### Acknowledgement

This research is being performed using funding received from the DOE Office of Nuclear Energy's Nuclear Energy University Program under contract DE-NE00009174. The authors would like to thank Dr. Sacit Cetiner at the INL and Dr. Joseph Lukens at the ORNL, for fruitful discussions and expert input.